# Regulated magnetic anisotropy and charge density wave in uniformly fabricated Janus CrTeSe monolayer

Jin-Hua Nie[1#], Cong Wang[2,3#], Mao-Peng Miao[1#], Kang-Di Niu[4#], Tao Xie[1], Ting-Fei Guo[1], Wen-Hao Zhang[1], Chao-Fei Liu[1], Rui-Jing Sun[1], Jian-Wang Zhou[1], Jun-Hao Lin[4], Wei Ji[2,3*] & Ying-Shuang Fu[1,5*]

1. School of Physics and Wuhan National High Magnetic Field Center, Huazhong University of Science and Technology, 430074 Wuhan, China
2. Beijing Key Laboratory of Optoelectronic Functional Materials and Micro-Nano Devices, Department of Physics, Renmin University of China, Beijing 100872, China
3. Key Laboratory of Quantum State Construction and Manipulation (Ministry of Education), Renmin University of China, Beijing 100872, China
4. Department of Physics, Southern University of Science and Technology, Shenzhen 518055, China
5. Wuhan Institute of Quantum Technology, Wuhan 430206, China

[#]These authors contribute to this work.
*Email: wji@ruc.edu.cn; yfu@hust.edu.cn

**Abstract**

**Two-dimensional materials with Janus structure host novel physical properties due to their inversional symmetry breaking. However, it remains elusive to synthesize Janus monolayer crystals with tailored long-range magnetic orders. Here, we have developed a general method to fabricate uniform Janus CrTeSe monolayers by selective selenization of preformed $CrTe_2$ monolayers with molecular beam epitaxy. The uniform Janus structure of CrTeSe with high crystal quality is confirmed by high-resolution scanning transmission electron microscopy. Spin-polarized scanning tunneling microscopy/spectroscopy measurements unveil that the Janus CrTeSe undergoes a charge density wave (CDW) transition and a**

**robust antiferromagnetic order. The magnetic anisotropy of CrTeSe is drastically altered compared to monolayer CrTe$_2$ by the breaking symmetries induced from the Janus structure and the CDW transition, as is substantiated with first principles calculations. Our research achieves the construction of large-area Janus structures, and artificially tailors the electronic and magnetic properties of Janus systems at the two-dimensional limit.**

**Introduction**

Two-dimensional (2D) Janus materials possess characteristic asymmetric sandwiched structures composed of distinct elements in their upper and lower layers. Such Janus structure violates central inversion symmetry of the system, bringing about a multitude of intriguing quantum phenomena, including giant second harmonic response[1,2], huge Rashba spin splitting[3], spontaneous electric polarization[1,4], as well as theoretically predicted intrinsic nonlinear planar Hall effect[5], and out-of-plane pyroelectric effect[6], etc.

The first successfully synthesized Janus monolayer crystal is MoSSe, which was obtained through atom substitution via CVD plasma and becomes now the primary means of obtaining Janus materials[1,7]. Several research groups have developed a variety of methods for preparing these materials (as summarized in Table S1), but the obtained Janus materials are still rare due to the cumbersome and complicated process[8-12]. Meanwhile, under the same chemical composition, except for rare examples such as the BiTeI family[3], the Janus structure is usually not the thermodynamically most stable phase under conventional conditions that tend to form hybrid alloys with multiple ratios[13], causing difficulty in synthesizing large-scale uniform Janus structures.

More importantly, the Janus material systems synthesized to date are largely limited to nonmagnetic transition metal chalcogenides. Magnetic Janus monolayer crystals, despite of their extensive theoretical investigations, predicting novel nonlinear spin textures[14-16], quantum anomalous Hall effect[17], etc., remains elusive in experiment. Therefore, it is under urgent demand to develop new ways of constructing Janus materials, particularly with magnetic order, to examine the predicted novel physics therein and offer material platforms for spintronic applications.

The system of choice is monolayer CrTeSe, because it is predicted to host noncollinear magnetism[18], and its parent compound monolayer $CrTe_2$ has been well-characterized previously[19]. In this study, we firstly achieved fabrication of *magnetic* Janus CrTeSe monolayers by selective selenization of preformed $CrTe_2$ monolayers with molecular beam epitaxy (MBE). The Janus CrTeSe has exceptionally uniform crystal quality, and undergoes a charge density wave (CDW) transition at low temperature. A zig-zag antiferromagnetic ground state is visualized with spin-polarized scanning tunneling microscopy (SP-STM). Magnetic anisotropy of the CrTeSe is drastically altered compared to its parent $CrTe_2$ monolayer both in magnitude and easy axis direction, as induced by the symmetry breaking effect from the Janus structure and the CDW transition.

**Result**

Our method of synthesizing Janus CrTeSe utilizes the different selenization conditions between the Te layers on the surface and the interface of monolayer $CrTe_2$. The monolayer $CrTe_2$ is grown on graphene-covered SiC substrate by coevaporating Te and Cr with MBE[19]. STM image of the as-grown monolayer $CrTe_2$ features a 2×1 stripe pattern (Fig. 1b), as reported in our previous work[19]. As an initial attempt, we follow similar elemental substitution procedure as Ref. 7, by annealing the preformed monolayer $CrTe_2$ under Se flux condition. Such method utilizes the different local Se pressure between the exposed top surface and the buried interface. As shown in Fig. S1, Janus CrTeSe indeed forms from the edges of monolayer $CrTe_2$ islands. However, extended annealing cannot transform the $CrTe_2$ homogenously into CrTeSe, because $CrSe_2$ already forms at partial regions before the whole surface transforms into CrTeSe, thus coexisting with $CrTe_2$. This demonstrates the local Se vapor pressure cannot be precisely controlled at atomic scale.

To avoid that issue, we develop an alternative Se capping method, whose flow chart is presented in Fig. 1a. Distinct to the Se flux method, the $CrTe_2$ films were firstly coated with Se layers up to ⩾500 nm thickness at room temperature, so that only the top Te layer directly interface the Se capping layers. Subsequently, the capped-$CrTe_2$ is annealed at a critical temperature of ~350 K, because lower annealing temperature results in inefficient selenization, and higher annealing temperature causes both top and bottom Te layers to selenize. STM image of the monolayer film after this procedure displays two different contrasts of Se and Te atoms on the top layer (Fig. 1c), where the Te atoms appear brighter due to their larger atomic radius, similar to FeTeSe[20] and MoSSe alloys[21]. Repeating the Se-capping/annealing procedures of about 5–10 cycles increases the ratio of surface Se atoms, until the whole surface is fully replaced with Se atoms (Fig. 1d). Generally, high coverage $CrTe_2$ requires lower cycles of Se-capping/annealing procedures to form the complete Janus CrTeSe structure (Fig. S2). Lattice constants the uniform Janus CrTeSe are measured as $a_1=a_2=3.4$ Å and $a_3=3.5$ Å, which are closer to those of $CrSe_2$[22], but smaller than $CrTe_2$ ($a_1=a_2=3.7$ Å, $a_3=3.4$ Å). Apparent height of monolayer CrTeSe is 740 pm, which is considerably smaller than that of $CrTe_2$ (800 pm) and similar to the Janus MoSSe[1]. The height reduction is due to the smaller Se atoms and shorter Cr-Se bond length[23]. We have further justified the universality of such Se capping method by to successfully preparing another Janus system of ZrTeSe (Fig. S3).

To confirm the synthesized Janus structure, we protect the CrTeSe film with thick Te capping layers and perform *ex-situ* scanning transmission electron microscopy (STEM) measurements. High-resolution cross-sectional STEM image reveals that the monolayer CrTeSe has identical 1T structure to that of the parent $CrTe_2$ (Fig. 1e). The top and bottom layer atoms show uniform darker and brighter contrast, respectively, without any signatures of intermixing. Such contrast reflects the different atomic numbers of the element[24], and thus corresponds to the lighter Se atoms and the heavier Te atoms, respectively. The Z-contrast imaging is nicely consistent with the simulated image (Fig. 1e). We simultaneously analyzed the ratio of chemical elements. The ratio of Cr:Se is much closer to 1:1, while that of Te is higher due to the Te capped layers (Figs. S4,S5). Line-scan electron energy loss spectroscopy of the cross-sectional CrTeSe monolayer shows a gradual change in the characteristic peaks along the Te-Cr-Se sequence (Fig. S6). These observations unambiguously validate the Janus structure of our CrTeSe film both in structure, chemical compositions, as well as its excellent uniformity.

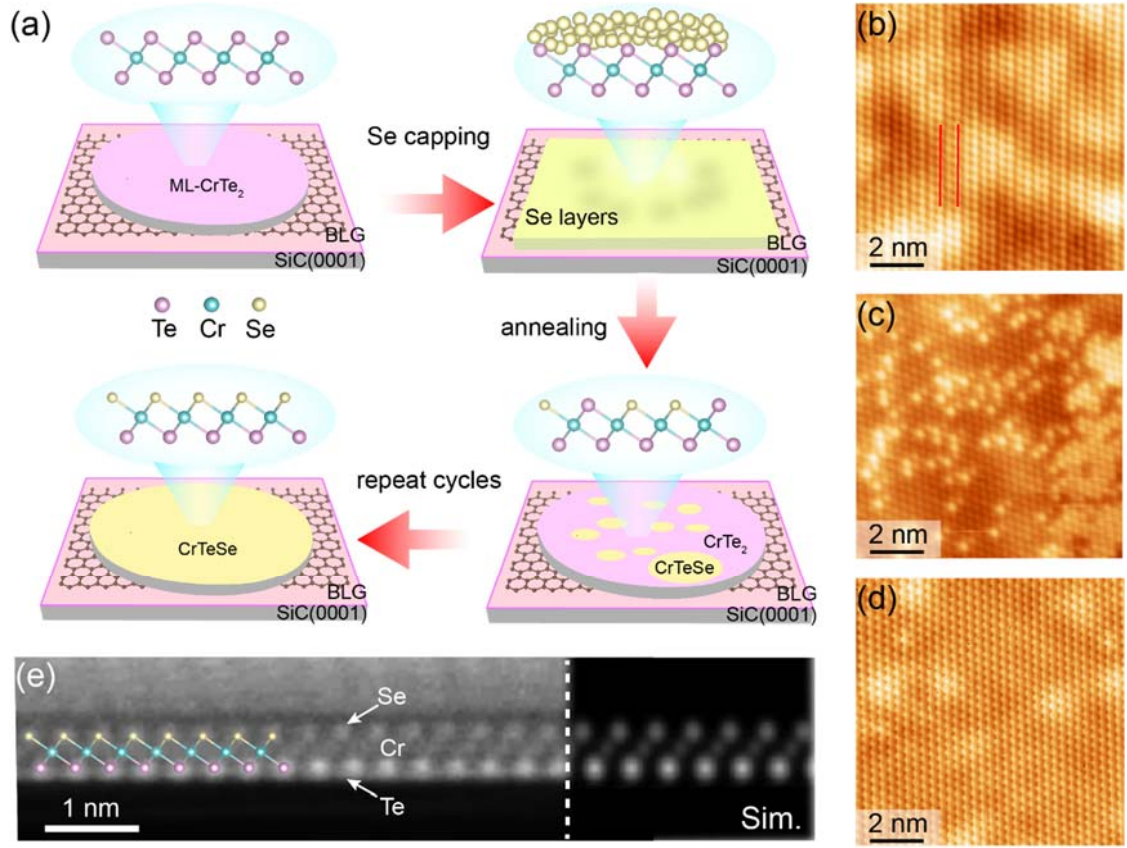

**Fig. 1 Schematic diagram and sample morphology for fabricating Janus CrTeSe.** (**a**) Flow chart of constructing Janus CrTeSe by selenizing monolayer $CrTe_2$ films. Step 1: growing ML-$CrTe_2$ on graphene-covered SiC substrate; Step 2: capping Se films on $CrTe_2$ at room temperature; Step 3: annealing at 350K to facilitate selenization; Step 4: repeating steps 2 and 3 to obtain uniform Janus CrTeSe. (**b-d**) Atom resolution STM images of $CrTe_2$, $CrTe_{1-x}Se_x$ and Janus CrTeSe, respectively. The red lines represent the typical 2×1 stripes of monolayer $CrTe_2$. Images condition: $V_b = 0.2$ V, $I_t = 100$ pA (**b**), $V_b = 0.2$ V, $I_t = 100$ pA (**c**), $V_b = 0.2$ V, $I_t = 100$ pA (**d**). (**e**) Cross-sectional STEM image of Janus CrTeSe and corresponding simulation. The contrast of top layer Se atoms is distinct from that of bottom layer Te atoms.

Having synthesized the high-quality Janus CrTeSe monolayer, we investigate its electronic properties. At 4 K, its STM image and associated fast Fourier transformation (FFT) image clearly resolve a $2\times\sqrt{3}$ superstructure (Fig. 2a), and the commonly observed 1×1 structure observed at 4 K becomes extremely rare. Electronic structure of the sample was measured with tunneling conductance spectra, which is proportional to its local density of states (LDOS). Monolayer $CrTe_2$ indicates an overall decreased electron density with increasing energy between -1.5 eV and 1 eV, with a spectral dip around the Fermi level (Fig. S7). In contrast to $CrTe_2$, the CrTeSe with the $2\times\sqrt{3}$ superstructure displays a pronounced peak at -0.73 eV and an evident gap of about 0.4 eV of finite conductance spanning the Fermi level. Such gap changes into a spectral dip for the CrTeSe in the 1×1 phase, and the pronounced peak also shifts to -0.68 eV.

To unravel the origin of the gap and the superstructure, we acquired a series of STM conductance mappings in constant height mode (Fig. S9). All the conductance mappings display the $2\times\sqrt{3}$ superstructure. Importantly, the conductance contrast of the superstructure exhibits an antiphase relation between the occupied (-0.3 eV) and unoccupied (0.1 eV) state around the gap edges (Figs. 2e,f). Such antiphase relation is also reflected in a 2D conductance plot acquired along a line traversing several periods of the $2\times\sqrt{3}$ superstructure (Fig. S9). The antiphase relation signifies the CDW origin of the gap and the superstructure[25].

A CDW state forms below a critical transition temperature. To further justify the CDW state, we investigated the evolution of conductance spectra and STM images at the same region of monolayer CrTeSe with temperature (Figs. S10 and S11). The CDW pattern prevails the whole sample surface below 60 K. With further elevating temperature, the area ratio of the 1×1 structure increases. After reaching 85 K, the whole surface transforms into the 1×1 structure. Similar temperature evolution occurs to the tunneling spectra, where the CDW gap gradually decreases with increasing temperature and finally closes simultaneously with the disappearance of the $2\times\sqrt{3}$ pattern.

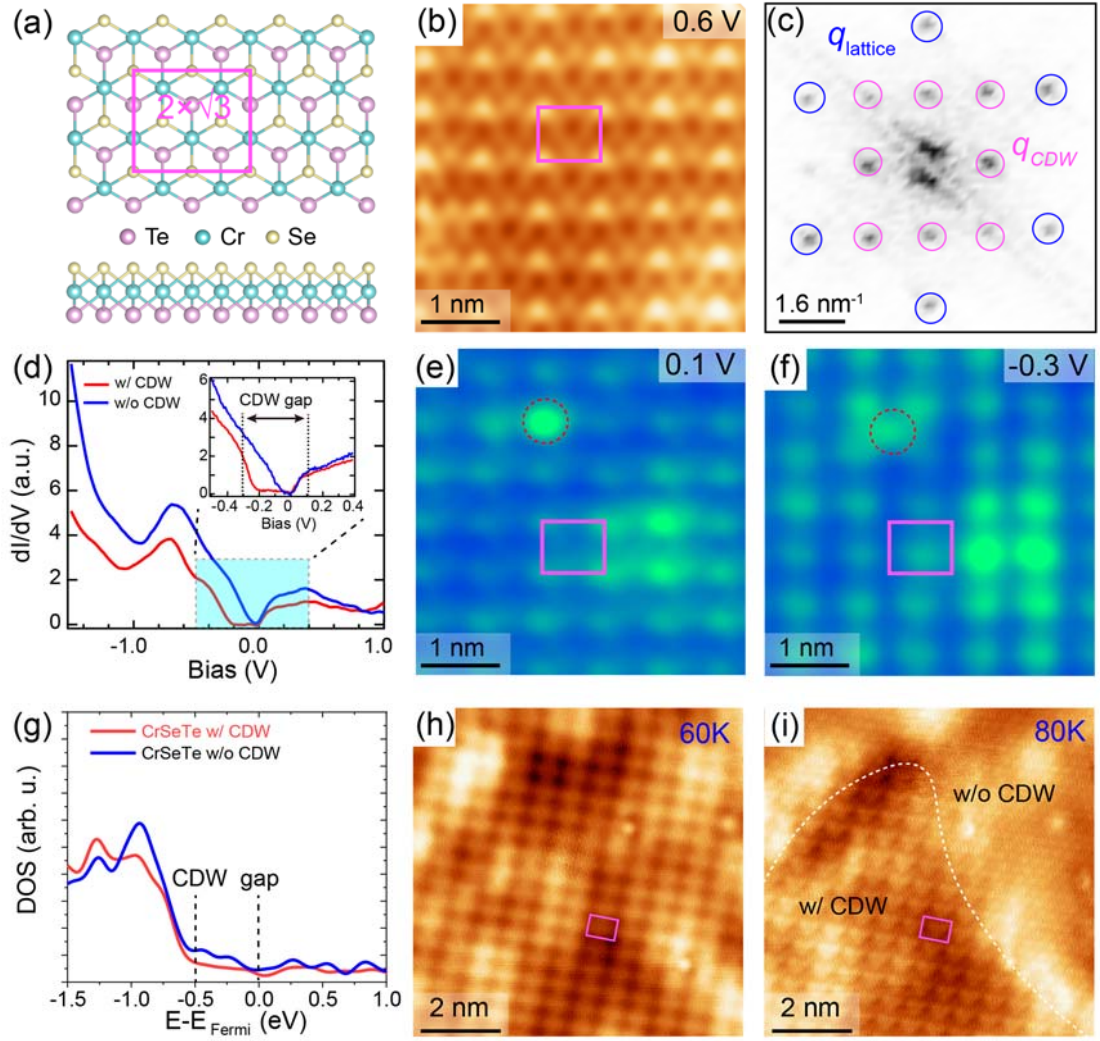

**Fig. 2 STM image and electronic structure of Janus CrTeSe at 4K.** (**a**) Top and side view of the atomic model of Janus CrTeSe. (**b**) The atomic resolution STM image ($I_t$ = 100 pA) of CrTeSe with the 2×√3 CDW pattern. The CDW unit is marked with magenta square in corresponding images. (**c**) The FFT image of **b**. The diffraction spots for the lattice and CDW are labled. (**d**) Differential conductance spectra ($V_b$ = 1.5 V, $I_t$ = 100 pA, $V_{mod}$ = 30 mV) for the CrTeSe with CDW (red curves) and without CDW (blue curves). Inset curves display the small range spectra ($V_b$ = 0.2 V, $I_t$ = 100 pA, $V_{mod}$ = 2 mV) around Fermi level, showing the CDW gap of ~0.4 eV on CrTeSe. (**e,f**) Constant height conductance mappings ($I_t$ = 100 pA, $V_{mod}$ = 20 mV), showing anti-phase relation between the occupied and unoccupied states around the CDW gap edges. A defect marker is labeled by red cycles. (**g**) Calculated DOS of monolayer Janus CrSeTe with and without CDW. A CDW gap of ~0.6 eV is identified, in reasonable agreement with the experiment. (**h,i**) STM images ($V_b$ = -0.5 V, $I_t$ = 30 pA) of the same region at different temperature, showing the evolution of the CDW pattern periods at 60 K (**h**) and 80 K (**i**). The phase boundary is exhibited by a white dotted line.

Apart from the CDW transition, the Janus structure may also alter the magnetization of CrTeSe, because its inversion symmetry breaking introduces DM interaction. Indeed, previous DFT calculations predicted a noncollinear spin order in monolayer CrTeSe due to the giant DM interaction[18]. However, the calculated lattice constant (3.56 Å) is much larger than the experimental value and no CDW is considered either, calling for an experimental determination of its magnetic order. SPSTM is a suitable probe in that regard. It is capable of resolving spin orders with atomic resolution via spin dependent electron tunneling, i.e. The tunneling current is dependent on the relative angle between the magnetization directions of the tip and sample. To determine the magnetic ground state of CrTeSe, we performed SPSTM measurements with an Fe-coated W tip, whose magnetization is in-plane due to its shape magnetic anisotropy and has the privilege of being tunable with magnetic field[26].

Figs. 3a-c and Fig. S12 show series of SPSTM images of monolayer CrTeSe under different magnetic fields. The CDW unit (white dotted boxe) serves as a marker since the Se atoms compromising the CDW unit always appear bright in each image. At 0 T, the two Se atoms marked with green segments in Fig. 3a appear bright, which are distinct to the image of Fig. 2b. One of the Se atom is located on the CDW unit. Upon application of a 1T field, two neighboring Se atoms (marked with green segments in Fig. 3c) become bright. Reversing the field to -1 T causes the intensity of Se atoms to recover back to the same contrast as that of 0 T. The contrast reversal under magnetic field demonstrates its origin from spin contrast. The atomic scale variation of spin contrast suggests the existence of AFM order in monolayer CrTeSe, similar to $CrTe_2$. We have also resolved similar spin contrast with an AFM Cr tip, whose magnetization is invariant against magnetic field. Notably, distinct to Figs. 3c,d, the spin contrast remains unaltered up to ±8 T (Fig.S13). This demonstrates the AFM state doesn't change under magnetic field, and the spin contrast reversal in Figs. 3c,d is from the rotation of the tip magnetization.

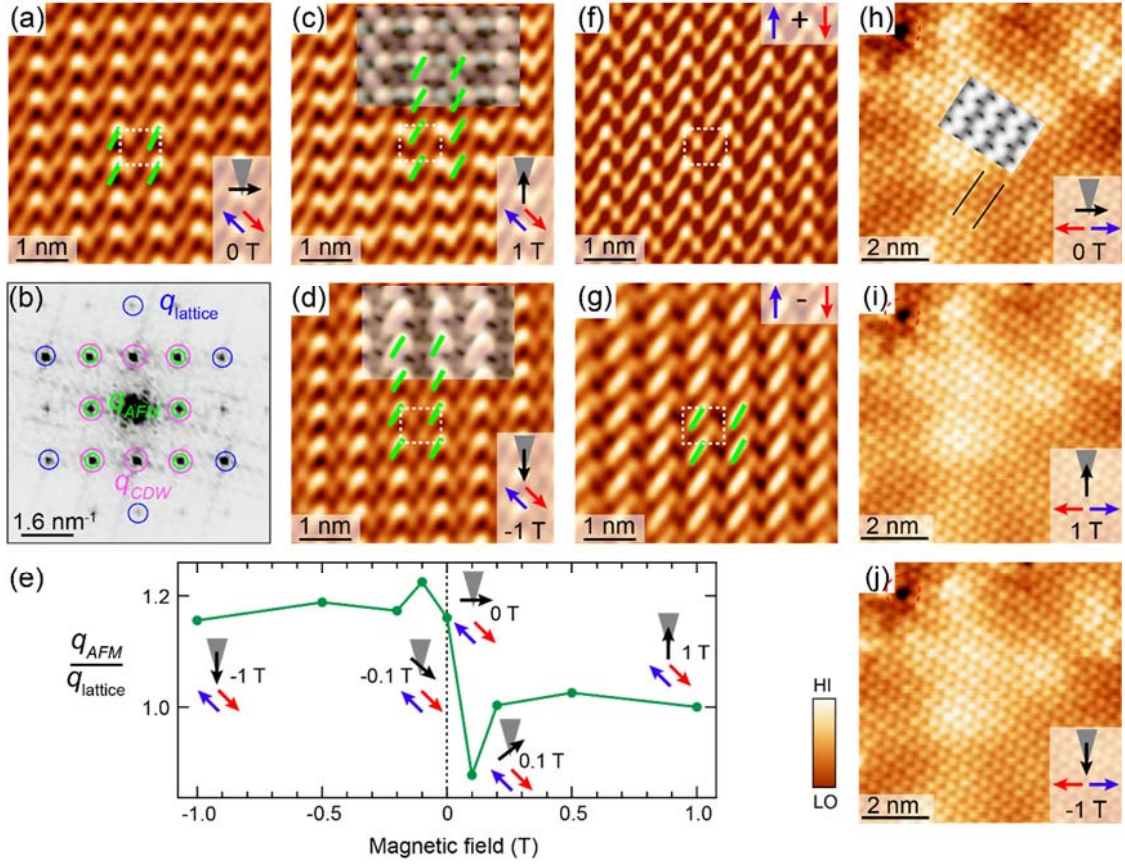

**Fig. 3 Spin mapping of the AFM order of Janus CrTeSe.** (**a,c,d**) SPSTM images ($V_b = 0.1$ V, $I_t = 100$ pA) of CrTeSe with an Fe tip at different magnetic field. The green segments highlight the Se atoms with bright spin contrast in corresponding images. A white dotted square represents the CDW unit. Inset images in (**c,d**) are simulated SPSTM images. (**b**) The FFT image of **a**. The CDW spots are marked with magenta circles and those of AFM are labeled by green circles. (**e**) The intensity ratio of $q_{AFM}/q_{lattice}$ with magnetic field. (**f,g**) The sum (**e**) and subtraction (**f**) of **b** and **c**, which exhibit the CDW periods and the AFM stripes, respectively. (**h-j**) SPSTM images ($V_b = 0.1$ V, $I_t = 100$ pA) of CrTeSe without CDW by using a Fe tip at different magnetic field. The black lines represent the zigzag AFM stripes, whose simulated SPSTM image is shown in the inset image. Magnetization configurations of the tip and sample are shown in corresponding images.

Next, we determine the magnetization configurations of the Fe tip and sample from the spin contrast evolution with magnetic fields. The spin contrast magnitude can be estimated from its fast Fourier transformation (FFT) of the SPSTM image. As is exemplified in Fig. 3b, diffraction spots of the CDW $q_{CDW}$ (magenta cycles) and AFM states $q_{AFM}$ (green circles) are superimposed. Nevertheless, the latter intensity significantly responses to the magnetic field, similar to those of monolayer CrTe$_2$[19]. The intensity of $q_{AFM}$ can be quantitatively analyzed from their comparison to the averaged intensity of the Bragg spots, $q_{lattice}$. As shown in Fig. 3e, such relative intensity $q_{AFM}/q_{lattice}$ stays constant upon the magnetic field exceeds 0.2 T for both of the field directions, since the tip magnetization is fully aligned along the field to out-of-plane. However, the $q_{AFM}/q_{lattice}$ value becomes slightly enhanced at -0.1 T, and obviously diminished at 0.1 T. This reflects the easy axis of the sample magnetization is titled, which is nearly in parallel (perpendicular) with the canted tip magnetization at -0.1 T (0.1 T), resulting the enhanced (diminished) spin contrast.

It is also noted that the $q_{AFM}/q_{lattice}$ value at negative field direction is ~17% larger than its positive counterpart, which is due to the superposition of the CDW pattern onto the AFM spin contrast at negative field. Such superposition can be separated by adding and subtracting the SPSTM images in Figs. 3c and d. This gives rise to the spin-averaged image of Fig. 3f that is similar to Fig. 2b, and the net spin contrast image of Fig. 3g that more clearly displays the zig-zag AFM spin order, respectively.

In monolayer CrTeSe, the Janus structure violates the inversion symmetry and the Peierls-type CDW pattern breaks translational symmetry, which substantially alter the magnetic anisotropy of its AFM order compared to that of monolayer $CrTe_2$. To disentangle the influence of the CDW pattern, we have further investigated the magnetic order of monolayer CrTeSe without CDW at 2 K. While nearly all the monolayer CrTeSe undergo the CDW transition, there are about 5% of the sample regions remain in the 1×1 phase (Supplementary Note 1). Those regions exhibit distinct zigzag stripes upon imaging with an Fe tip at 0 T (Fig. 3h), which disappear at ±1T (Figs. 3i,j), suggesting their origin from a zigzag AFM order with in-plane magnetic anisotropy.

To reveal the stabilization mechanism of the metastable Janus structures, monolayer CrSeTe in hexagonal structures with Se and Te atoms occupying different sites (Fig. S14) were considered in our density functional theory (DFT) calculations. Formation enthalpies of the CrSeTe monolayer configurations are calculated using the total energy method (refer to the Methods). Our calculations show negative formation enthalpies of CrSeTe monolayers in the Te rich and deficient limit, which correspond to the spontaneous selenization transition (Fig. 4a). In the freestanding case, the non-Janus configurations with Se and Te atoms coexisting in the same sublayer (Fig. S15) is more stable than the Janus structure by 32 meV/Cr. However, after considering the substrate induced interfacial epitaxial strain, the Janus structure becomes the ground state and is at least 22 meV/Cr more stable than other configurations. In addition, the formation enthalpy of the Janus structure is significantly lower than that of the miscibility of $CrTe_2$ and $CrSe_2$, ruling out the possibility of phase separation.

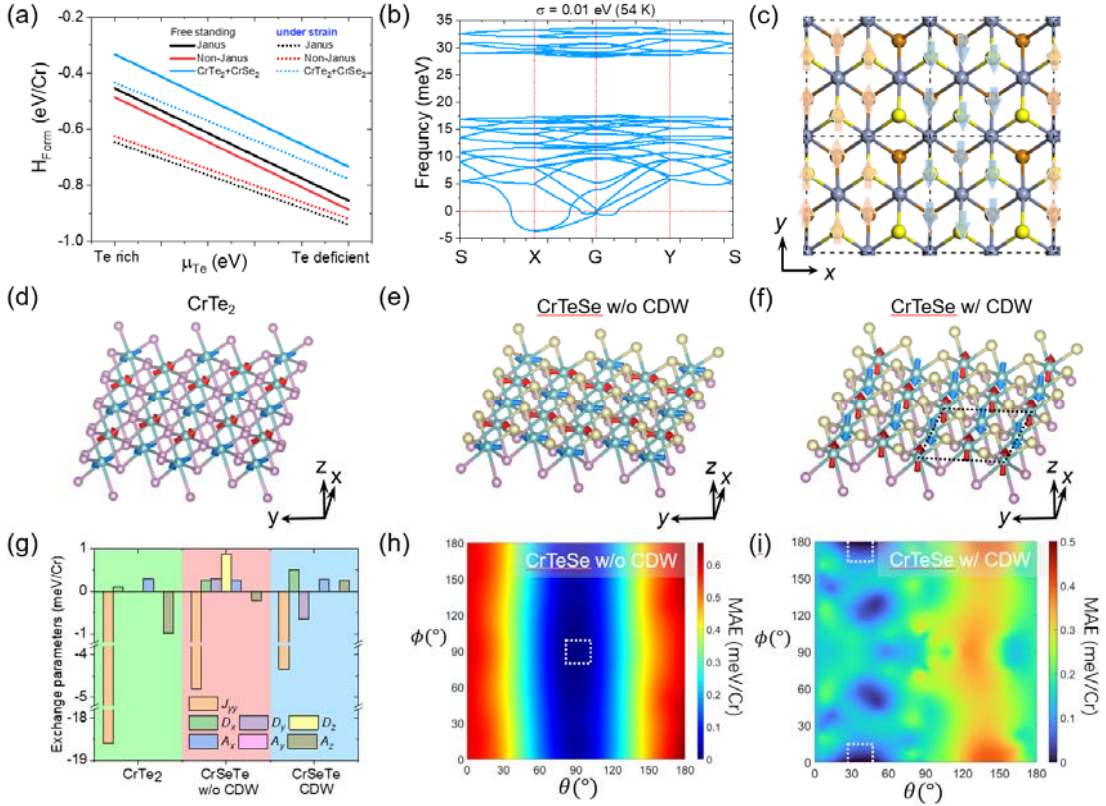

**Fig. 4. CDW phase transition induced spin reorientation in Janus CrSeTe monolayer.** (**a**) Enthalpies of formation for the CrSeTe monolayer with Se and Te atoms occupying different sites. (**b**) Phonon dispersion of 1×1 Janus CrSeTe at a smearing factor of 54 K using the Fermi-Dirac function as smearing function. (**c**) Schematic of 2×√3 CDW structure of Janus CrSeTe monolayer. The slate-blue, yellow, and orange balls represent Cr, top Se, and bottom Te atoms, respectively. Salmon pink and light blue arrows correspond to the direction of atomic displacement after CDW transition. (**d-f**) Calculated zigzag AFM groundstate and easy axis direction of ML $CrTe_2$ (**d**), Janus CrSeTe without (**e**) and with CDW (**f**). Green and red arrows represent the magnetization directions of Cr atoms. (**g**) Exchange parameters of pristine monolayer $CrTe_2$, Janus CrSeTe with and without CDW. (**h-i**) Angular dependence of the calculated MAE of Janus CrSeTe without (**h**) and with CDW (**i**). Here, θ and ϕ correspond to the angles between the magnetization direction and the z and x axis, respectively. A step size of 10° is used in our calculations. The total energy of the Cr moment oriented to the easy axis direction was chosen as the zero-energy reference.

Subsequently, we calculated the phonon spectrum of the Janus CrSeTe monolayer with 1×1 structure (Fig. 4b). The phonon spectrum exhibits significant softening at point X (0.5, 0, 0), combined with the zigzag AFM order to form the 2×√3 CDW period. We thus constructed the CDW phase based on the vibration displacements of modes with imaginary frequency and found it is 1.4 meV/Cr more stable compared with the pristine CrSeTe (Fig. 4c). The stability of the CDW phase is also supported by the absence of imaginary values in the calculated vibrational frequencies at a smearing factor of 54 K. Moreover, calculated DOS of the Janus CrSeTe both with and without the CDW state also all reasonably agree with the experiment (Fig. 2g).

The zigzag order is the magnetic ground state in both $CrSe_2$ and $CrTe_2$ monolayer[19,27,28], which remains unchanged after the formation of Janus structure (Fig. 4d-f, Table S2) with the corresponding simulated spin-resolved image based on the zigzag AFM configuration well consistent with our SPSTM images (Fig. 3). Our calculation found the introduction of the Janus structure first shifts the orientation of the easy axis originally tilted in $CrTe_2$ monolayer (Fig. 4d) to the *y* axis direction (Fig. 4e and 4h). Further CDW phase transition reorientates the easy axis to a titled orientation 47° off the *z*-axis (Fig. 4f and 4i), consistent with our experiments (Fig. 3). To unveil the manipulative mechanism of the easy axis direction, we calculated the nearest anisotropic exchange parameter J, DMI parameters and single ion anisotropic parameters based on an anisotropic Heisenberg model (Fig.4g and Fig. S18). Introducing of the Janus structure significantly influences the electron band structure around the Fermi level and thereby weaken the contribution of Te atoms, leading to four times weakened single ion anisotropic energy $A_z$ (Fig. S18). Although Janus induced DMI is still not larger enough compared to J to change the zigzag magnetic ground state, the strongest $D_z$ causes the magnetic moment to tend to follow the in xy-plane direction (Fig. 4h). After the CDW phase transition, the in-plane symmetry breaking further leads to enhanced electric dipole along the *z* axis and significantly enhances the $D_y$ (Fig. S18), leading to the magnetic moment to point in a titled orientation 47° off the *z*-axis (Fig. 4i).

In conclusion, we achieved the fabrication of large-scale Janus CrTeSe monolayer crystals by developing a general strategy of in-situ selenizing the top Te layer of preformed monolayer $CrTe_2$ with MBE. Using SPSTM, we observed that the Janus CrTeSe undergoes a charge density wave (CDW) transition at ~90 K and a zig-zag antiferromagnetic ground state. More importantly, both the magnitude and the orientation of the magnetic anisotropy are drastically altered compared to its parent $CrTe_2$ monolayer. Such observations are substantiated with DFT calculations, unveiling that the Janus structure and the CDW transition brings about symmetry breaking, which enhances the direct AFM exchange interaction. Our developed method of constructing large-area Janus structures envisions to be generalized to other 2D materials, as has been demonstrated in monolayer ZrTeSe. The Janus structure opens up a versatile platform for artificially tailoring the physical properties of materials at the monolayer limit.

**Note added:** We noticed a related study on Janus VTeSe by Ziqiang Xu *et al.* (arXiv:2406.12180) posted during the preparation of this manuscript. While that study also observed a CDW transition, the CDW pattern is different to ours and no magnetic order was reported in Janus VTeSe either.

**Acknowledgement:** We gratefully acknowledge the financial support from the National Key Research and Development Program of China (Grant No. 2022YFA1402400), the National Natural Science Foundation of China (Grants Nos. 92265201, U20A6002, 11974422, 12104504, 12204534, and 12174131), the Ministry of Science and Technology of China (Grant No. 2023YFA1406500, Grant No. 2018YFE0202700), the Strategic Priority Research Program of Chinese Academy of Sciences (Grant No. XDB30000000), the Fundamental Research Funds for the Central

Universities, the Research Funds of Renmin University of China (Grants No. 22XNKJ30) (W.J.). TEM characterizations received assistance from SUSTech Core Research Facilities and technical support from Pico Creative Centre, which is also duly supported by the Presidential Fund and Development and Reform Commission of Shenzhen Municipality. All calculations for this study were performed at the Physics Lab of High-Performance Computing (PLHPC) and the Public Computing Cloud (PCC) of Renmin University of China.

**Author contributions**

J.H.N. and M.P.M. grew the sample and did the SPSTM experiments with the help of X.T., T.F.G., W.H.Z., C.F.L., R.J.S. and J.W.Z.; K.D.N. and J.H.L. carried out the STEM experiments; C.W. and W.J. performed the calculations; Y.S.F., W.J., J.H.N., C.W., M.P.M. and K.D.N. analysed the data, and wrote the manuscript, with comments from all authors. Y.S.F. and W.J. supervised the project.

**Methods**

**Molecular-beam epitaxy (MBE) growth.** The graphene substrate was obtained via graphitisation of the SiC(0001) substrate with cycles of vacuum flashing treatment, the details of which are depicted in Ref. 29. The graphene obtained with such a treatment is dominated with bilayers. The 1T-$CrTe_2$ films were grown on the graphene-covered SiC(0001) substrate by the same conditions in Ref. 19. For the Se flux method of selenizing $CrTe_2$, the Se flux was kept at 20nm/min. The annealing temperature of CrTeSe with Se capped layers was lower than 370 K. For *ex-situ* measurements, the $CrTe_2$ films were protected against degradation with Te capping layers of ~200 nm in thickness.

**STM measurements.** The measurements were performed in a custom-made Unisoku STM system mainly at 4.2 K unless described exclusively. The spin-averaged STM data were measured with an electrochemically etched W wire, which had been characterised on a Ag(111) surface prior to the measurements. The SP-STM data were taken with Cr or Fe tips. The Cr tip was prepared by coating ~50 layers of Cr (purity: 99.995%) on a W tip, which had been flashed to ~2000 K to remove oxides, followed by annealing at ~500 K for 10 min. The Fe tip was prepared by coating ~30 layers of Fe on a W tip, following similar flash and annealing procedures as the Cr tip. The tunnelling spectra were obtained through a lock-in detection of the tunnelling current with a modulation voltage of 983 Hz. The topographic images were processed by WSxM.

**STEM imaging.** A cross-sectional STEM sample was prepared using a focused ion beam. To ensure structural integrity, the surface of Te-capping CrTeSe was covered with graphite through a routine dry transfer method. STEM imaging and EDS analysis were conducted on an FEI Titan Themis instrument, which is equipped with an X-FEG electron gun and a DCOR aberration corrector, operating at 300 kV. The inner and outer collection angles ($\beta1$ and $\beta2$) were set at 48 mrad and 200 mrad, respectively. The convergent half-angle was set at 25 mrad and the electron beam current was approximately 60 pA. All imaging procedures were performed at ambient room temperature.

**DFT calculations.** Our density functional theory (DFT) calculations were performed using the generalized gradient approximation (GGA) for the exchange correlation potential in the form of PerdewBurke–Ernzerhof (PBE)[30], the projector augmented wave method[31], and a plane-wave basis set as implemented in the Vienna ab-initio simulation package (VASP)[32,33]. Dispersion correction was considered in Grimme's semiempirical D3 scheme[34] in combination with the PBE functional (PBE-D3). This combination achieves accuracy comparable to that of the optB86b-vdW functional for describing geometric properties of layered materials at a lower computational cost[35]. Kinetic energy cutoffs of 700 and 600 eV for the planewave basis were adopted for structural relaxations and electronic structure calculations, respectively. All atoms, lattice volumes, and shapes were allowed to relax until the residual force per atom was less than 0.01 eV/Å. A vacuum layer of over 15 Å in thickness was used to reduce imaging interactions between adjacent supercells. A Gamma-centered $k$-mesh of 21×21×1 was used to sample the first Brillouin zone of the unit cell for monolayer

MnSe$_2$. A Gaussian smearing method with a $\sigma$ value of 0.01 eV was used for all calculations. The on-site Coulomb interaction to Cr $d$ orbitals was characterized by $U$ and $J$ values of 3.0 eV and 0.6 eV[19], as determined via a linear response method[36] and validated by comparing theoretical prediction and experimental results. A $2\times 2\sqrt{3}$ supercell and four (eight) magnetic configurations (Fig. S19) were considered to find the magnetic ground state for the monolayer (bilayer). Formation enthalpies of the CrSeTe monolayer configurations are calculated using the total energy method and is defined as: $\Delta H_{Form} = E_{CrSeTe} - E_{CrTe_2} + \mu_{Te} - \mu_{Se}$, where $\mu_{Te}$ and $\mu_{Se}$ are the chemical potential of the removed Te and added Se atoms to form the CrSeTe monolayer. Chemical potentials of Cr and Te in CrTe$_2$ fulfill the equation $\mu_{Cr} + 2\mu_{Te} = \mu_{Cr}^* + 2\mu_{Te}^* + \Delta H_{CrTe_2}$, where $\mu_{Cr}^*$ and $\mu_{Te}^*$ correspond to the chemical potential of Cr and Te in the most stable bulk form, $\Delta H_{CrTe_2}$ is the formation enthalpy of CrTe$_2$. We could thus deduce the range of $\mu_{Te}$ as $\mu_{Te}^* + \frac{1}{2}\Delta H_{CrTe_2} < \mu_{Te} < \mu_{Te}^*$. While the chemical potential of Se is considered in the rich limit by choosing the most stable bulk form.